\documentclass[aps,showpacs,twocolumn,nobibnotes]{revtex4}
\usepackage{graphics,graphicx,amsfonts,amsmath,amsbsy,amssymb,color}
\usepackage{bm}
\usepackage{subfigure}
\usepackage{threeparttable}

\begin{document}

\title{%
Explicitly correlated plane waves: Accelerating convergence in periodic wavefunction expansions
}

\author{Andreas~Gr\"uneis$^{(a,b)}$}
\email{andreas.grueneis@univie.ac.at}
\author{James~J.~Shepherd$^{(a)}$}
\author{Ali~Alavi$^{(a)}$}
\author{David~P.~Tew$^{(c)}$}
\author{George~H.~Booth$^{(a)}$}
\email{ghb24@cam.ac.uk}
\affiliation{%
$^{(a)}$University of Cambridge, Chemistry Department,
Lensfield Road, Cambridge CB2 1EQ, U.K.}
\affiliation{%
$^{(b)}$Faculty of Physics and Center for Computational Materials Science, 
University Vienna, Sensengasse 8/12, A-1090 Vienna, Austria.}
\affiliation{$^{(c)}$School of Chemistry, University of Bristol, Bristol BS8 1TS, U.K.}

\begin{abstract}
We present an investigation into the use of an explicitly correlated plane wave basis for periodic wavefunction expansions at
the level of second-order M\o ller-Plesset perturbation theory (MP2).
The convergence of the electronic correlation energy with respect to the
one-electron basis set is investigated and compared to conventional
MP2 theory in a 
finite homogeneous electron gas model.
In addition to the widely used Slater-type geminal correlation factor, we also derive and
investigate a novel correlation factor that we term Yukawa-Coulomb.
The Yukawa-Coulomb correlation factor  is motivated
by analytic results for two electrons in a box and
allows for a further improved convergence
of the correlation energies with respect to the employed basis set.
We find the combination of the infinitely delocalized plane waves and local short-ranged geminals provides
a complementary, and rapidly convergent basis for the description of periodic wavefunctions.
We hope that this approach will expand the scope of discrete wavefunction expansions in periodic systems.
\end{abstract}
\date{\today}

\pacs{}

\keywords{}
\maketitle

\section{\label{sec:introduction}Introduction}

Despite huge advantages in terms of accuracy and systematic improvability, wavefunction-based quantum chemical methods are routinely used
by only a small fraction of electronic structure theorists, in contrast to density functional theory (DFT) which
dominates the community\cite{Weitao2012}. Nowhere is this more true than in the solid state, where application of high-level quantum chemistry
methods are only beginning to emerge in a recently growing field
\cite{BGKA2013,Scuseria2001,Stoll2005,Schutz2007,Pisani2007,Kresse2009,Manby2009,Kresse2010,Gillan2010,Manby2010,Kresse2011,Schutz2011,Stoll2012,Paulus2012,Vandevondele2012,Manby2012}.
The reason for this slow uptake is the computational cost of these
methods, which generally scale as a high power of the system size, compared to the lower mean-field scaling of DFT. This is exacerbated in the
solid state where increasing the size of the supercell to converge finite size effects is far more costly than for mean-field counterparts. Much of this
expense originates from the need to expand out the many-electron wavefunction in terms of anti-symmetrized one-particle functions of a
specified basis set. This itself must then be expanded and generally extrapolated to near completeness to obtain 
accurate results and justify the use of the high level of correlation treatment. 
Although methods more familiar to the solid state, such as DFT\cite{Burke2013} and Diffusion Monte Carlo (DMC)\cite{Foulkes2001} require a basis
set, there is only a weak dependence since no many-electron wavefunctions are expanded in this basis.

The difficulty with the expansion of many-electron wavefunctions as antisymmetric products of one-particle basis functions (Slater determinants) has been known 
since the early days of electronic structure theory, and is due to the short-ranged or `dynamic' correlation between 
electron pairs. As the electrons coalesce, a derivative discontinuity or `cusp' must arise, so that a divergence in the
kinetic energy operator cancels an opposite one in the potential. Within an expansion of Slater determinants the exact cusp is never obtained, and
a quantitatively correct linear form at small interelectronic distances only arises with large basis sets of high momenta. 
A description of these cusps was initially formulated by Kato\cite{Kato:CPAM10-151,Pack:JCP45-556,Ashcroft82,Sahni2003}, who 
found the wavefunction to be linear to first-order as a function of the interelectronic distance between the pairs ($r_{12}$). Moreover, the gradient of
this linear behavior was found to be exactly a half (or quarter for triplet pairs), regardless of the form of the rest of the potential 
in the system. Higher order terms in $r_{12}$ however are affected by the rest of the potential\cite{Tew08}.

For many years, methods were developed which tried to exploit this knowledge of the form of the exact wavefunction in the small $r_{12}$ limit,
but the methods which resulted, such as methods utilizing exponentially correlated Gaussians\cite{Boys1960} and the 
transcorrelated method\cite{Handy69} among others, were generally expensive, plagued
by many electron integrals, and limited to systems of only a small size. A major breakthrough was achieved in 1985 by Kutzelnigg, where two electron geminal
functions were introduced into the wavefunction which satisfied the electron cusps, and augmented a traditional Slater 
determinant expansion\cite{Kutzelnigg:TCA68-445,Klopper87,Kutzelnigg:JCP94-1985}. This resulting 
wavefunction expansion was then used within the formulation of traditional quantum chemical methods, and crucially, an approximate resolution of 
identity (RI) was performed as a way to factorize the many-electron integrals into sums of products of at most two-electron quantities. A small set of these 
geminals dramatically improved the convergence of quantum chemical methods with respect to basis set size, since fewer 
high momenta functions were required for these energetically significant cusp regions. This dual basis of traditional determinants and strongly orthogonal
geminals, and the methods for evaluating resultant expectation values, has been named F12 theory.

In the intervening years this approach has matured, with important advances taking it from a promising technique to an 
indispensable tool for high-accuracy quantum chemical methods for large 
systems\cite{Hattig:CR112-4,Kong:CR112-75,Tew:BOOK2010,Werner:BOOK2010,Tenno12,Tenno12_2,Shiozaki2010}. 
These advances include the introduction of a complementary auxiliary basis set in which to perform 
the RI\cite{Klopper:JCP116-6397,Valeev:CPL395-190}, refinement of the approximations used in order to minimize the impact of the RI and 
maintain orbital invariance\cite{Kedzuch:IJQC105-929,Ten-no:JCP121-117,Werner:JCP126-164102,Tew:MP108-315,Manby05}, a more general function of the interelectronic 
coordinate to approximately capture longer range effects\cite{Ten-no:CPL398-56,Tew:JCP123-074101}, and the introduction 
of specially designed basis sets for optimal efficiency\cite{Peterson08,Tew09}. The result are methods 
which share the intrinsic accuracy of the complete basis set (CBS) limit of their
parent method, but which approach this limit far more rapidly, thereby reducing the cost of the method. Combining this with 
density fitting\cite{Manby03}, local approximations\cite{Tew11,Werner11}, and multireference methods\cite{Gdanitz:CPL210-253,Ten-no:CPL446-175,Shiozaki:JCP131-141103,Shiozaki:JCP134-034113,Shiozaki:JCP134-184104,Kedzuch:CPL511-418,Haunschild:CPL531-247,Torheyden:JCP131-171103,Kong:JCP135-214105,Kong:JCP133-174126,Booth2012,Werner13} has greatly extended the
reach of quantum chemistry in recent years.

All F12 approaches to date have taken place within the framework of a traditional atom-centered Gaussian basis set. Although these
functions are ubiquitous in gas-phase molecular quantum chemistry, where their local nature generally suits the wavefunction,
it is unclear whether these are well suited for extended systems, especially when the wavefunction is intrinsically delocalized. These systems have been 
traditionally studied in a discrete basis of plane waves, chosen such that the boundary conditions at the edges of the unit cell are fulfilled,
although this is by no means the only choice in solids. 
However, a plane wave basis confers many advantages in the solid state. There is a single
basis set parameter (the orbital kinetic energy cutoff), which allows the CBS limit to be approached systematically and straightforwardly, without
the need for basis set optimization. These basis functions are also strictly orthogonal, and therefore no issues with linear dependencies 
occur as the basis increases, in contrast to Gaussian functions.

However, for all these advantages of a plane wave basis, the features of electronic cusps are still
missing, and are difficult to capture without including very high energy plane waves
in the expansion which dramatically increases the cost. This convergence has been found to have the same scaling behavior as the Gaussian 
expansion\cite{Shepherd2012_3,Kutzelnigg92}, though generally requires many more functions to reach the complete basis limit.

In this paper, we attempt to overcome these difficulties by combining a plane wave basis with the explicitly correlated F12 approach,
and evaluate energies at the level 
of second-order M\o ller--Plesset (MP) theory to analyze the benefit.
We first consider the 3D finite-electron uniform electron gas (UEG) for this approach,
which has recently received attention as a model system for wavefunction-based quantum
chemistry~\cite{Ashcroft82,Grueneis2009,Shepherd2012_1,Shepherd2012_2,Shepherd2012_3,Shepherd2013,Pederiva2013}, as well as long 
being an important model, especially in the development of density functional theory\cite{Ceperley1980,Perdew1981}.
As the simplest model for a fully-periodic metallic system, it has many advantages.
The plane waves in the UEG are exact natural orbitals, but in addition they
are also exact Hartree--Fock solutions, and kinetic energy eigenfunctions. This means that 
the generalized Brillouin condition (GBC) and the
extended Brillouin condition (EBC) are exactly satisfied, which decouples the conventional and F12 energy 
contributions\cite{Klopper87,Klopper:JCP116-6397,Manby05}. 

In addition, all three-electron integrals
have simple analytic forms, whose RI can be saturated completely with the addition of at most just a single auxiliary orbital.
Tractable expressions for the electron repulsion integrals mean that extrapolation to the CBS limit
is straightforward to derive and understand; these energies can be easily found and used as
benchmarks~\cite{Shepherd2012_3}. We note in passing that the CBS limit is also well-defined
for the MP2 energy of a finite system, even though the energy diverges in the thermodynamic
limit~\cite{Kresse2010,Shepherd2013}. This is because the divergence is caused by low-momenta
excitations in the large box limit, rather than the high-momenta basis functions responsible for converging the basis set incompleteness error.

The simple model Hamiltonian also allows us to calculate the
exact MP1 wavefunction for the two electron UEG analytically, whose expansion about $r_{12} =0$ we find to take a different form than the traditional Slater-type correlation factor now established 
in molecular F12 theory. We use this to compare the Slater-type form to a new correlation factor which we find to be optimal for the UEG, and which may have 
advantages in other solid-state (or even potentially molecular) systems.
Finally, we apply the method to the most
widely studied solid-state system with quantum chemical methods, rocksalt lithium hydride crystal, to
check the transferability of the findings into realistic {\em ab initio} solid state systems.

\section{\label{sec:theory}Theory}
This section outlines the theoretical methods that are employed in the present work to
study the uniform electron gas simulation cell Hamiltonian.
We briefly recapitulate second-order M\o ller-Plesset perturbation (MP2) theory,
explicit correlation and the Hylleraas functional.
Furthermore, we elaborate on the use of a plane-wave basis set in the many-electron wavefunction
expansion and its implications for explicitly correlated methods.
Analytical expressions for the integrals required in the
above methods are derived and techniques to treat finite size effects as well as
singularities are discussed.
Finally a new correlation factor that we term Yukawa-Coulomb correlation factor is derived.

\subsection{\label{ssec:mp2}Second-order M\o ller-Plesset perturbation theory}

In MP2 theory, electron correlation is treated using many-body Rayleigh-Schr\"odinger perturbation
theory, taking the $N$-electron Fock operator as the unperturbed Hamiltonian $H^{(0)}$ \cite{Moller}.
The Hartree--Fock wave function $|\Psi^{(0)} \rangle$ both defines and is defined by the
Fock operator. Formally, it is the ground state Slater determinant with occupied orbitals that are eigenstates of 
the $1$-electron Fock operator
\begin{equation}
F \vert i \rangle  = \epsilon_i \vert i \rangle 
\end{equation}
In practical computations, however, the $\vert i \rangle$ are rarely true eigenstates of the 
Fock operator, since they are expressed in a finite and in general, insufficient $1$-electron basis. 
Nevertheless, these Hartree--Fock orbitals define $H^{(0)}$. For the UEG, which is 
the main focus of this work, the Hartree--Fock orbitals are determined by symmetry. They are
therefore exact eigenstates and the generalized and extended Brillouin conditions are fulfilled.
For the UEG, $|\Psi^{(0)} \rangle$ is the exact ground state of the zeroth-order Hamiltonian $H^{(0)}$.

In MP2 theory the standard route to obtaining the first-order wavefunction is to expand it in the 
basis of excited Slater determinants $|\Psi_{ij}^{ab}\rangle$:
\begin{equation}
|\Psi^{(1)}\rangle = \frac{1}{2}\sum_{ij}^{\rm occ.}\sum_{ab}^{\rm virt.} t_{ij}^{ab} E_{ij}^{ab}|\Psi^{(0)} \rangle,
\label{eq:psi1}
\end{equation}
where $i$, $j$ and $a$, $b$ refer to occupied and unoccupied spatial Hartree--Fock orbitals respectively, from the full set of
$M$ one-electron basis functions. $E_{ij}^{ab}$ is the spin-free two electron excitation operator.
The coefficients of the excited determinants $t_{ij}^{ab}$ are readily calculated and read
\begin{equation}
t_{ij}^{ab}=\frac{\langle ij | ab \rangle}{\epsilon_i + \epsilon_j -\epsilon_a -\epsilon_b}.
\end{equation}
In the above expression, $\epsilon_n$ corresponds to the one-electron HF eigenvalues, and $\langle ij | ab \rangle$
are the conventional electron repulsion integrals. 
The calculation of two-electron integrals will be outlined in Sec.~\ref{ssec:integralevaluation}.
In M\o ller-Plesset perturbation theory
the second-order energy is the leading order correction to the correlation energy that can be obtained
by calculating $\langle\Psi^{(0)}|H-H^{(0)}|\Psi^{(1)}\rangle$, which simplifies to
\begin{equation}
E_c^{\rm MP2}=\sum_{ij}^{\rm occ.}\sum_{ab}^{\rm virt.} \frac{\langle ij | ab \rangle(
2\langle ab | ij \rangle - \langle ba | ij \rangle)}
{\epsilon_i + \epsilon_j -\epsilon_a -\epsilon_b}.
\end{equation}
Both the energy and the equations that determine the first-order wavefunction separate into decoupled equations for
each occupied pair. 
The pair correlation energy can alternatively be obtained by optimizing the first-order pair correlation function $\vert u_{ij}\rangle$
to minimize the Hylleraas energy functional
\begin{equation}
E_c^{ij} = \min [ \langle u_{ij} | F_1 + F_2 - \epsilon_i - \epsilon_j | u_{ij} \rangle + 2 \langle u_{ij} | \frac{1}{r_{12}} | ij \rangle ].
\label{eq:hylleraas}
\end{equation}
This expression is useful in explicitly correlated methods. In conventional MP2 theory, the spinless first-order pair function and its
contravariant counterpart, have the expansion
\begin{align}
| u_{ij} \rangle &= \frac{1}{2} \sum_{ab}^{\rm virt.} t_{ij}^{ab} | ab \rangle, \\
\langle  u_{ij} \vert &= \frac{1}{2} \sum_{ab}^{\rm virt.}  \langle ab |(2t_{ij}^{ab} - t_{ij}^{ba}).
\label{eq:uij}
\end{align}

\begin{figure}
\includegraphics[width=8.0cm,clip=true]{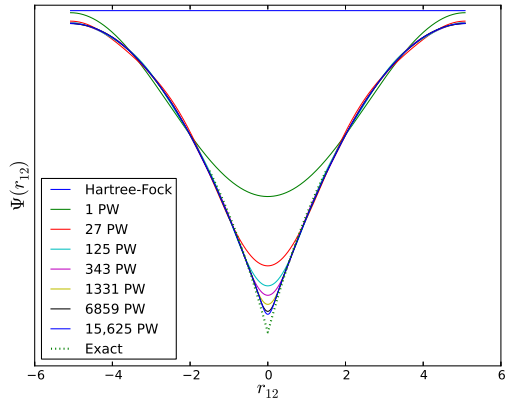}
\caption{\label{fig:psitwoelec} 
The MP1 wavefunction for the two electron uniform electron gas at $r_s = 5$~a.u. with increasing plane wave orbital basis sets, up
to a total basis of 15,625 plane waves. 
One electron is fixed at the center of the box, and the other is moved in a line through the coalescence point. This
illustrates the slow convergence to the exact wavefunction as the interelectronic distance tends to zero. 
The Hartree--Fock wavefunction shows no variation with interelectronic distance, as only the average 
electronic potential is felt across the box. This demonstrates the same qualitative cusp convergence 
in plane waves as demonstrated elsewhere for molecular systems in Gaussian basis sets\cite{Hattig:CR112-4,Kong:CR112-75}.
}
\end{figure}

Figure~\ref{fig:psitwoelec} visualizes the zeroth- (HF) and first-order wavefunctions using the example of two electrons in a box with
a homogeneous neutralizing background charge. The wavefunctions are plotted with respect to the interelectronic
distance $r_{12}$.
While the first-order wavefunction accounts for electronic correlation by decreasing the probability of finding both
electrons at short interelectronic distances, the zeroth-order wavefunction does not exhibit this so-called correlation hole 
centered at the electron coalescence point ($r_{12}$=0), and is depicted by a flat line.
We note that the first-order wavefunction converges very slowly to a cusp in the complete basis set limit with respect
to the number of employed virtual one-electron orbitals used in the expansion of $| \Psi^{(1)} \rangle$ [see Eq.~(\ref{eq:psi1})].

\subsection{\label{ssec:f12}Explicitly correlated MP2}

As outlined above and in Refs.~\onlinecite{Shepherd2012_3,Hattig:CR112-4,Kong:CR112-75}, the MP1 wavefunction converges frustratingly slowly
to the complete basis set limit, with an $M^{-1}$ dependence.
Concomitantly the correlation energy converges very slowly and usually requires the treatment of a large
one-electron basis sets of high momenta, that result in significant computational effort.
However, as shown in figure~\ref{fig:psitwoelec}, a large fraction of the basis is needed to describe the many-electron
wavefunction in the vicinity of the electron-electron coalescence points.
The first-order cusp condition defines the
shape of the many-electron wavefunction close to the electron coalescence of singlet pairs~\cite{Kato:CPAM10-151}
\begin{equation}
\left . \frac{\partial \left ( \frac{\Psi({\bf r}_{ij})}{\Psi(0)} \right ) }{\partial  {\bf r}_{ij}} \right |_{{\bf r}_{ij}=0}=\frac{1}{2} \quad .
\end{equation}
The above equation implies that the many-electron wavefunction exhibits a 
derivative discontinuity about ${\bf{r}}_{12}=0$, with a linear behavior as you
move from this point regardless of the external potential of the system, and as seen in figure~\ref{fig:psitwoelec}.
Explicitly correlated methods fulfill the first-order cusp conditions exactly by augmenting the ansatz
for the many-electron wavefunction with two-electron terms that depend on the interelectronic distance
of each electron pair. In explicitly correlated MP2-F12 theory\cite{Werner:JCP126-164102,Bachorz:JCC32-2492}
the first-order pair functions $| u_{ij}^{\rm F12} \rangle$ are expanded as
\begin{equation}
| u_{ij}^{\rm F12} \rangle = \frac{1}{2} \sum_{ab}^{\rm virt.} t_{ij}^{ab} | ab \rangle + t_{ij} \hat{Q}_{12} f_{12} \vert ij \rangle,
\label{eq:psif12}
\end{equation}
where $t_{ij}$ are geminal amplitudes determined by the universal cusp conditions,
$f_{12}$ is the correlation factor that models the shape of
the correlation hole and is typically chosen to be a Slater-type function\cite{Ten-no:CPL398-56}
\begin{equation}
f_{12}^{\rm STG}= e^{-\gamma r_{12}} .  \label{eq:STGCorrFac}
\end{equation}
This choice ensures that the geminal functions included in the basis are linear with respect to $r_{12}$ in the vicinity of the electron-electron
cusp and decay to zero at large $r_{12}$, where the wavefunction is expected to vary smoothly and is generally well-represented by the conventional
determinantal basis. We note that it is more common in explicitly correlated Gaussian implementations to approximate this 
functional form by a fixed combination of Gaussian type correlation factors to simplify integral evaluation over this kernel, however this is not a problem 
here (see section~\ref{ssec:integralevaluation}), and an exact Slater-type geminal function is used. 

This form of the correlation factor is an empirical choice, and its longer-ranged decay is not motivated 
by an underlying theory, but rather intuition\cite{Ten-no:CPL398-56}. However, it has been shown to be accurate compared to various other alternatives 
in molecular systems\cite{Tew:JCP123-074101,Tew:JCP125-094302}.
In molecules, it is likely that the rapid exponential decay of the correlation factor to zero (the lack of long range structure)
is an advantage because it separates out the long-range behavior which is not expected to be able to be modeled by a simple function of $r_{12}$ due to the
anisotropy of the external potential\cite{Tew:JCP125-094302}. 
Furthermore, in Ref.~\onlinecite{Tew:JCP123-074101} the function has been
shown to not increase monotonically to a constant, but
rather to reach a maximum and then decrease, due to the fact that the remaining molecular electron density 
is reduced at large interelectronic distances.

However here we will also investigate a new correlation factor derived from perturbation theory that we term Yukawa-Coulomb correlation factor,
\begin{equation}
f_{12}^{\rm YC}= \frac{2}{\gamma}\frac{1-e^{-\gamma r_{12}}}{r_{12}}.
\label{eq:yc}
\end{equation}
The MP2-F12 theory outlined in this work is, however, independent from the specific form of $f_{12}$.
Therefore we will return to the discussion of $f_{12}^{\rm YC}$ in Sec.~\ref{sec:yukcoulcorr}.

The projector $\hat{Q}_{12}$ enforces strong orthogonality between $|\Psi^{(1)} \rangle$ and $|\Psi^{(0)} \rangle$,
and it also enforces orthogonality between the standard and F12 contributions to the first-order wave function.
\begin{equation}
\hat{Q}_{12}=(1-\hat{O}_1)(1-\hat{O}_2)-\hat{V}_1\hat{V}_2,
\label{eq:proj}
\end{equation}
where
\begin{equation*}
O_1=\sum_i^{\rm occ.} |r'_1 \rangle \langle r'_1 | i \rangle \langle i | r_1 \rangle \langle r_1 | , { ~~~~}
V_1=\sum_a^{\rm virt.}|r'_1 \rangle \langle r'_1 | a \rangle \langle a | r_1 \rangle \langle r_1 |
\end{equation*}

In F12 theory it is convenient to obtain the second-order correlation energy by optimizing $| u_{ij}^{\rm F12} \rangle$
to minimize the Hylleraas functional Eq.~(\ref{eq:hylleraas}).
The F12 contributions involve non-factorizable many-electron integrals, which increase the computational cost of
evaluating $E_c^{\rm MP2-F12}$ compared to $E_c^{\rm MP2}$.
The calculation of many-electron integrals
can, however, be approximated by the introduction of resolutions of identity using the orbital basis, and an additional 
orthogonal complimentary auxiliary basis set (CABS).
In this work we employ unoccupied Hartree--Fock orbitals as CABS. 

For calculations on the UEG, since the EBC is fulfilled, the contributions involving $t_{ij}^{ab}$ do not depend on the
F12 terms\cite{Klopper87,Manby05,Hattig:CR112-4}. The energy then decomposes into the standard MP2 correlation energy and
an F12 correction
\begin{equation}
E_c^{\rm MP2-F12}(M) =E_c^{\rm MP2}(M) + E_c^{\rm F12}(M).
\end{equation}
The dependence of the above energies on $M$ indicates that these energies
are calculated using a finite orbital basis that is composed of
$M$ plane waves with a kinetic energy below a specified cutoff.
The complete basis set limit is approached for $M\rightarrow\infty$.
The expressions for $E_c^{\rm F12}$ have been derived elsewhere\cite{Werner:JCP126-164102,Hattig:CR112-4,Kong:CR112-75,Shiozaki2010}, and are given here as
\begin{align}
 E_c^{\rm F12}(M)=&2 V_{mn}^{ij} ( 2 t_{mn}^{ij} - t_{nm}^{ij} ) \nonumber  \\
  & + t_{mn}^{kl} B_{kl}^{ij} ( 2 t_{mn}^{ij} - t_{nm}^{ij} ) \nonumber  \\
  & - (\epsilon_m + \epsilon_n) t_{mn}^{kl} X_{kl}^{ij} (2 t_{mn}^{ij} - t_{nm}^{ij}).
\label{eq:ecf12}
\end{align}
In the above expression, the indices $i$, $j$, $k$, $l$, $m$ and $n$ refer to occupied HF orbitals, and Einstein summation convention is assumed.
$t_{kl}^{ij}$ are the geminal amplitudes that fulfill the first-order cusp condition and are kept
fixed at the diagonal orbital-invariant ansatz of Ten-no\cite{Ten-no:JCP121-117}, which exactly satisfy the first order
cusp-conditions of singlet and triplet electron pairs,
\begin{align}
t_{ii}^{ii}=-\frac{1}{2}\gamma^{-1} \label{eq:tii}\\
t_{ij}^{ij}=-\frac{3}{8}\gamma^{-1} \label{eq:tij}\\
t_{ij}^{ji}=-\frac{1}{8}\gamma^{-1} \label{eq:tji}.
\end{align}

\begin{table}[t]
\caption{
Index notation for different orbital subspaces of the complete one-electron basis. 
CABS refers to the complimentary auxiliary basis set\cite{Valeev:CPL395-190}, and OBS refers to the 
orbital basis set over which the conventional MP1 amplitudes are defined.
}
\label{tab:index}
\begin{ruledtabular}
\begin{tabular}{lccc}
                        &  Occ. OBS orbitals & Virt. OBS orbitals & CABS \\ \hline
$i,j,k,l,m,n$           &  Yes       &  No  & No \\
$a,b$                   &   No       &  Yes & No \\
$p,q$                   &  Yes       &  Yes & No \\
$P,Q,R$                 &  Yes       &  Yes & Yes \\
$a'$                    &   No       &   No & Yes \\
\end{tabular}
\end{ruledtabular}
\end{table}

The intermediates $V$, $X$ and $B$ are defined as, 
\begin{align}
V_{mn}^{ij}=&Y_{mn}^{ij}-R_{mn}^{pq} v_{pq}^{ij}-R_{mn}^{la'} v_{la'}^{ij}-R_{mn}^{a'l} v_{a'l}^{ij} \label{eq:vintdef}\\
X_{mn}^{ij}=&\bar{R}_{mn}^{ij}-R_{mn}^{pq} R_{pq}^{ij}-R_{mn}^{la'} R_{la'}^{ij}-R_{mn}^{a'l} R_{a'l}^{ij} \label{eq:xintdef} \\
B_{mn}^{ij}=&{\tau}_{mn}^{ij} + \hat{S}_{12} \left (\frac{1}{2} \hat{S}_{H} \bar{R}_{mP}^{ij} h_n^P \right . \nonumber\\
&-R_{mn}^{PQ} k_P^R R_{RQ}^{ij} - R_{mn}^{Pk} f_P^Q R_{Qk}^{ij} \nonumber \\
&+R_{mn}^{ka'} f_k^l R_{la'}^{ij} - R_{mn}^{pa} f_p^q R_{qa}^{ij} \nonumber \\
&\left . -\hat{S}_{H} R_{mn}^{ka'} f_k^P R_{Pa'}^{ij} -\hat{S}_{H} R_{mn}^{a'b} f_{a'}^p R_{pb}^{ij} \right )
\end{align}
For this work the $B$ intermediate is calculated using approximation C~\cite{Kedzuch:IJQC105-929,Noga2008}.
Table~\ref{tab:index} summarizes the meaning of the above indices.
$f_P^Q$, $h_P^Q$ and $k_P^Q$ refer to the Fock-, Hartree- and exchange matrix. We note
that $f_P^Q=h_P^Q-k_P^Q$. The following section outlines their evaluation for the UEG.
$Y_{mn}^{ij}$, $R_{mn}^{ij}$, $\bar{R}_{mn}^{ij}$, ${\tau}_{mn}^{ij}$ and $v_{mn}^{ij}$
correspond to two electron integrals defined as
\begin{align}
Y_{mn}^{ij}=&      \left \langle \phi_m \phi_n  \left | f_{12}v_{12} \right | \phi_i \phi_j \right \rangle \label{eq:yukawar} \\
R_{mn}^{ij}=&      \left \langle \phi_m \phi_n  \left | f_{12} \right | \phi_i \phi_j \right \rangle  \label{eq:f12r} \\
\bar{R}_{mn}^{ij}=&\left \langle \phi_m \phi_n  \left | {f}_{12}^2 \right | \phi_i \phi_j \right \rangle \label{eq:squf12r} \\
{\tau}_{mn}^{ij}=&\left \langle \phi_m \phi_n  \left | (\nabla_1 f_{12})^2 \right | \phi_i \phi_j \right \rangle \label{eq:taur} \\
v_{mn}^{ij}=&      \left \langle \phi_m \phi_n  \left | v_{12} \right | \phi_i \phi_j \right \rangle.
\label{eq:coulr}
\end{align}
$f_{12}$ and $v_{12}$ is the correlation factor and the electron repulsion kernel, respectively.
We will return to the evaluation of the above integrals in reciprocal space in Sec.~\ref{ssec:integralevaluation}, which is performed
in the Vienna ab-initio simulation package ({\tt VASP})\cite{Kresse96}.
The operators $\hat{S}_{12}$ and $\hat{S}_{H}$ symmetrize four index quantities such that
\begin{equation}
\hat{S}_{12} T_{mn}^{ij}= T_{mn}^{ij} + T_{nm}^{ji}.
\end{equation}
\begin{equation}
\hat{S}_{H} T_{mn}^{ij}= T_{mn}^{ij} + T_{ij}^{mn}.
\end{equation}
We stress that the above expressions hold for general systems with real as well as
complex electron repulsion integrals, so that the introduction of $k$-point symmetry in {\em ab initio} systems follows naturally.

\subsection{\label{ssec:pwbasis}The homogeneous electron gas in a plane wave basis set}

In this work we seek to apply explicitly correlated second-order M\o ller-Plesset perturbation theory
to a finite-size (insulating) uniform electron gas model.
The $N$-electron homogeneous electron gas simulation-cell Hamiltonian reads
\begin{equation}
\hat{H}=-\sum_\alpha \frac{1}{2} \nabla^2_\alpha + \sum_{\alpha, \beta} \frac{1}{2}\hat{v}_{\alpha \beta},
\end{equation}
where $\alpha$ and $\beta$ are electron indices and the two-electron Ewald interaction $\hat{v}_{\alpha \beta}$
is given by
\begin{equation}
\hat{v}_{\alpha \beta}=\frac{1}{\Omega} \sum_{{\bf G}} \frac{4\pi}{{\bf G}^2}e^{i {\bf G} ({\bf r}_\alpha-{\bf r}_\beta)} ,
\end{equation}
and $\Omega$ refers to the volume of the real-space simulation cell.
For all calculations in the present work, we employ a cubic real-space unit cell
with 54 electrons unless stated otherwise.
The reciprocal lattice vectors $\bf G$ are defined as
\begin{equation}
{\bf G}=\frac{2 \pi}{L} \left( \begin{array}{c} n \\ m \\ l \end{array} \right)
\end{equation}
where $n$, $m$, and $l$ are integer numbers and $L$ is the real-space box length
such that $L^3=\Omega$.
The one-electron orbitals are chosen to be plane waves
\begin{equation}
\phi_n({\bf r})=\frac{1}{\sqrt{\Omega}} e^{i {{\bf k}_{n} {\bf r}}},
\label{eq:pworb}
\end{equation}
where $\bf k$ refers to the unique reciprocal lattice vector of the orbital.
The one-electron Hartree--Fock Hamiltonian becomes diagonal in this orbital basis and reads
\begin{align}
\left \langle \phi_n \left | H^{(0)} \right | \phi_m \right \rangle =& f_n^m = \delta_{n,m} (h_n^m-k_n^m)=\epsilon_n, {~~\rm where} \nonumber \\
h_n^n=&\frac{1}{2}{\bf k_n}^2 {~~\rm and} \\
k_n^n=&- \sum_i \left \langle ni \left | v_{12} \right | in \right \rangle. \label{eq:exchange}
\end{align}

\subsection{\label{ssec:integralevaluation}Evaluation of the integrals in reciprocal space}

It can be advantageous to calculate the electron repulsion integrals
in reciprocal space if a plane-wave basis set is employed.
This reduces the computational effort from a six-dimensional integral in real space
to a three-dimensional sum in reciprocal space over the Fourier components of the given electron pair codensities
\begin{align}
\langle ij | v_{12} | ab \rangle=\sum_{\bf G} C_{ia{\bf G}}  \tilde{{v}}_{\bf G}  C^*_{bj \bf G},
\label{eq:erirec}
\end{align}
where
\begin{equation}
\sum_{{\bf G}} C_{ia \bf G} e^{i {\bf G} {\bf r}}=\phi_i^*({\bf r})\phi_a({\bf r}).
\label{eq:olapden}
\end{equation}
The Fourier components
of the integral kernels
in Eqs.~(\ref{eq:f12r}) and (\ref{eq:coulr}) read
\begin{align}
\tilde{f}_{\bf G}^{\rm STG}=\mathcal{FT}\left ( f_{12}^{\rm STG} \right )=&\frac{4 \pi}{({\bf G}^2+\gamma^2)^2}   \label{eq:f12} \\
\tilde{f}_{\bf G}^{\rm YC}=\mathcal{FT}\left ( f_{12}^{\rm YC} \right )=&\frac{4 \pi}{({\bf G}^2+\gamma^2){\bf G}^2}.
\label{eq:corryc} \\
\tilde{v}_{\bf G}=\mathcal{FT}\left ( v_{12} \right )=&\frac{4 \pi}{{\bf G}^2} \label{eq:potcoul} 
\end{align}
We note that if the orbitals correspond to plane waves, as it is the case in the UEG, momentum conservation applies.
$\langle ij | v_{12} | ab \rangle$ is non-zero only if ${\bf k}_i+{\bf k}_j={\bf k}_a+{\bf k}_b$.
Moreover, in the UEG all orbital codensities, and therefore two-electron integrals can be defined uniquely 
from the momentum transfer vector ${\bf k}_i-{\bf k}_a$ such that
\begin{equation}
\langle ij | v_{12} | ab \rangle=\tilde{{v}}_{{\bf k}_i-{\bf k}_a}.
\label{eq:uegeri}
\end{equation}

\subsubsection{\label{ssec:singularities}Treatment of singularities in reciprocal potentials}

The reciprocal kernels in Equations~(\ref{eq:corryc}) and (\ref{eq:potcoul}) diverge at ${\bf G}=0$.
Although these singularities become only problematic for integrals $\langle vw | vw \rangle$ (due to the orthogonality of the orbitals),
a direct numerical evaluation of the ${\bf G}=0$ contribution to the electron repulsion integrals
according to Eq.~(\ref{eq:uegeri}) is not possible.
The singularities are, however, integrable and well-known solutions to this problem
have already been proposed~\cite{Gygi1986}.
We will employ a technique that introduces a Gaussian charge distribution $C_{\bf G}$ whose integral over the reciprocal space
with the corresponding kernels can be calculated analytically as
\begin{equation}
\frac{1}{\Omega}
\sum_{\bf G} C_{\bf G} \tilde{v}_{\bf G} \rightarrow \int d{\bf G} e^{-\alpha {\bf G}^2} \tilde{v}_{\bf G},{~~\rm where~} C_{\bf G}= e^{-\alpha {\bf G}^2}.
\end{equation}
$\alpha$ is chosen such that the charge distribution decays to zero at the
boundary of the employed plane wave grid and is constant in the vicinity of ${\bf G}=0$.
Adding and removing this Gaussian charge distribution to $C_{ia{\bf G}}C^*_{bj{\bf G}}$ on the right-hand side of Eq.~(\ref{eq:erirec}) gives
\begin{align}
&\frac{1}{\Omega}\sum_{{\bf G}} (C_{nn{\bf G}}C^*_{mm{\bf G}}-C_{\bf G}+C_{\bf G}) \tilde{{v}}_{\bf G} \nonumber \\
=&\frac{1}{\Omega}\underbrace{\sum_{{\bf G}} (C_{nn{\bf G}}C^*_{mm{\bf G}}-C_{\bf G})\tilde{{v}}_{\bf G}}_{{\bf G}=0~{\rm contribution~vanishes}}
+\frac{1}{\Omega}\underbrace{\sum_{{\bf G}} C_{\bf G}\tilde{{v}}_{\bf G}}_{\rm analytical~integration}.
\label{eq:erirec2}
\end{align}
The difference between the Gaussian and orbital charge distribution vanishes for ${\bf G}=0$,
removing the ${\bf G}=0$ contribution from the sum in the first term on the right-hand side
of the above equation.
The last term on the right-hand side can be integrated analytically.
Depending on the kernel, we obtain the following results for the integrals.
\begin{widetext}
\begin{align}
\frac{1}{\Omega} \sum_{{\bf G}} C_{\bf G}\tilde{{v}}_{\bf G} \rightarrow \int d{\bf G} \frac{4 \pi e^{-\alpha {\bf G}^2} }{{\bf G}^2}=& 2\pi \sqrt{\frac{\pi}{\alpha}}  \\
\frac{1}{\Omega} \sum_{{\bf G}} C_{\bf G}\tilde{{f}}^{\rm YC}_{\bf G} \rightarrow \int d{\bf G} \frac{4 \pi e^{-\alpha {\bf G}^2} }{({\bf G}^2+\gamma^2){\bf G}^2}=&
\frac{2 \pi ^2 e^{\alpha  \gamma ^2} \text{Erfc}\left(\sqrt{\alpha } \gamma \right)}{\gamma }. 
\end{align}
\end{widetext}
Practically speaking, the ${\bf G}=0$ component is computed once per kernel and stored.
This one-time effort does not require significant optimization.

\subsubsection{\label{ssec:conv}Convolution of integral kernels in reciprocal space}

We compute the reciprocal kernels for the integrals in Eqs.~(\ref{eq:yukawar}), (\ref{eq:squf12r}) and (\ref{eq:taur})
using the convolution theorem with
\begin{align}
\mathcal{FT}\left ( f_{12}v_{12} \right )  &=\frac{1}{\Omega}\sum_{{\bf G}'} \tilde{v}_{{\bf G-G}'} \tilde{f}_{{\bf G}'} \label{eq:yukawaconv} \\
\mathcal{FT}\left ( {f}_{12}^2 \right ) &=\frac{1}{\Omega}\sum_{{\bf G}'} \tilde{f}_{{\bf G-G}'} \tilde{f}_{{\bf G}'} \label{eq:squf12conv} \\
\mathcal{FT}\left ( (\nabla_1 f_{12})^2 \right )    &=\frac{1}{\Omega}\sum_{{\bf G}'} \tilde{f}_{{\bf G-G}'} \tilde{f}_{{\bf G}'}({\bf G}\cdot{\bf G}' - {\bf G}' \cdot {\bf G}')  \label{eq:tauconv}. 
\end{align}
The integral kernels are calculated using the convolution theorem in order to 
treat finite-size effects in the $B$ intermediate consistently and
obtain the correct limiting behavior for the F12 contributions away from the large box-size limit.
We stress that $E_c^{\rm F12}$ must vanish in the complete basis set limit ($M\rightarrow\infty$) in a non-trivial way, 
since the conventional determinant amplitudes recover the CBS energy in this limit.
Specifically, the contributions of the $V$, $X$ and $B$ intermediates to $E_c^{\rm F12}$ must all
vanish individually.
In the following we will discuss this behavior for the $V$ intermediate that reads 
\begin{equation}
V_{mn}^{ij}=Y_{mn}^{ij}-R_{mn}^{pq} v_{pq}^{ij}-R_{mn}^{la'} v_{la'}^{ij}-R_{mn}^{a'l} v_{a'l}^{ij}.
\end{equation}
The first term in the above equation on the right hand side must cancel
with the others as the employed basis set approaches completeness.
The contraction over the orbital indices in the last three terms corresponds to a 
resolution of identity between the Coulomb potential (present in the $v_{pq}^{ij}$ integrals) and the correlation 
factor (present in the $R_{mn}^{pq}$ integrals).
Thus it is important that the three different integral kernels ($\frac{1}{r}$, $e^{-\gamma r}$
and $\frac{e^{-\gamma r}}{r}$) are treated in a consistent manner, which is achieved via the convolution theorem.


\subsection{\label{sec:yukcoulcorr}A new MP2-F12 correlation factor: Yukawa-Coulomb}

The optimal correlation factor maximizes the convergence rate
of the MP2-F12 correlation energy to the CBS limit with respect to the employed orbital basis set.
All MP2-F12 implementations have so far been confined to molecular systems,
where different choices of correlation factors have been investigated but did not yield
an improvement over the conventional Slater-type correlation factor\cite{Tew:JCP123-074101}.
In this work we seek to investigate a correlation factor 
motivated by analytic results for two electrons in a box with a neutralizing and uniform background charge\cite{Gaskell61,Ceperley78,Ashcroft82}.
The amplitudes of the first-order wavefunction for two electrons with opposite spins in a box are given by
\begin{equation}
t_{ii}^{ab}=\frac{\langle ii | ab \rangle}{\epsilon_i + \epsilon_i -\epsilon_a -\epsilon_b}, \label{eq:mp1amplitudes}
\end{equation}
where $\vert i \rangle = \Omega^{-1/2}$ is the spatial orbital at the gamma point ${\bf G}=0$.
In this case the kinetic energy of the occupied orbitals are zero and
momentum conservation of all two-electron excitations requires that ${\bf k}_b=-{\bf k}_a$. Therefore the 
denominator of Eq.~(\ref{eq:mp1amplitudes}) can be approximated by
\begin{equation}
\epsilon_i + \epsilon_i -\epsilon_a -\epsilon_b \approx -{\bf k}_a^2+ \tilde{\gamma}. \label{eq:denomapprox}
\end{equation}
In the above equation, we have approximated the contributions of the exchange $k_a^a$ [see equation~(\ref{eq:exchange})]
to the HF one-electron energies by a constant $\tilde{\gamma}$. We note that
in the limit ${\bf k}_a \rightarrow \infty$, the denominator will be dominated by contributions
of the kinetic energy whereas the exchange contributions
to $\epsilon_a$ will decay as $1/{\bf k}_a^2$.
Inserting the definition of the electron repulsion integrals and noting that ${\bf k}_a$ in this instance is also equal to the momentum transfer 
vector of the excitation, the above approximation 
gives
\begin{equation}
t_{ii}^{ab}=-\frac{1}{\Omega} \frac{4 \pi}{{\bf k}_a^2 ({\bf k}_a^2 - \tilde{\gamma})}.
\end{equation}
A sum over all orbital products in the reciprocal lattice to obtain the wavefunction form then allows for an analytic inverse Fourier transform 
of the electron pair function to real space, to yield the first-order pair function
\begin{equation}
\vert u_{ii} \rangle = - \frac{2}{\gamma^2}\frac{1-e^{-\gamma r_{12}}}{r_{12}} \frac{1}{\Omega}
\end{equation}
with $\gamma^2 = \tilde \gamma$.
The corresponding correlation factor consistent with Eq.~(\ref{eq:psif12}) and Eq.~(\ref{eq:tii}) is
\begin{equation}
f^{\rm YC}_{12}=\frac{2}{\gamma}\frac{1-e^{-\gamma r_{12}}}{r_{12}}. \label{eq:yccorrf}
\end{equation}
The above correlation factor, that we denote Yukawa-Coulomb correlation factor,
becomes linear in $r_{12}$ for $r_{12}\rightarrow0$ and decays to zero for large $r_{12}$.
We note that the Yukawa-Coulomb correlation factor is similar to the two-body Jastrow factor used
in previous studies of the homogeneous electron gas with transcorrelated methods\cite{Umezawa2004}.
This correlation factor may equivalently be derived directly in real space, starting from the differential equation
for the first-order wave function for doubly occupied pairs in a UEG,
\begin{align}
(F_1 + F_2 - \epsilon_i - \epsilon_j) Q_{12} f(r_{12}) \frac{e^{i 2{{\bf k}_{i} {\bf s}}} }{\Omega} + Q_{12} \frac{1}{r_{12}} \frac{e^{i 2{{\bf k}_{i} {\bf s}}}}{ \Omega} = 0
\end{align}
where ${\bf s} = ({\bf r}_1 + {\bf r}_2)/2$ and we have asserted that the first-order pair function can be exactly 
represented by the product of the $ij$ orbital pair with an isotropic function of $r_{12}$. Since the GBC and
EBC are fulfilled, $[Q_{12},F_1]=0$ and we can therefore solve for $f(r_{12})$ without considering $Q_{12}$. 
Approximating $F_1 + F_2 \approx T_1 + T_2 + \tilde \gamma = -\nabla^2_{r_{12}} - \frac{1}{4} \nabla^2_s + \tilde \gamma$ gives
\begin{align}
(-\nabla^2_{r_{12}} + k_i^2 - \epsilon_i - \epsilon_j + \tilde \gamma)  f(r_{12}) + \frac{1}{r_{12}} = 0
\end{align}
which has the solution $f(r_{12}) = -f_{12}^{\rm YC}/2\gamma$ with $\gamma^2 = k_i^2 - 2\epsilon_i + \tilde \gamma$.

The main difference between Eq.~(\ref{eq:yccorrf}) and the Slater-type function is in the longer-ranged behavior, as $f^{\rm YC}$
decays to zero as $1/r_{12}$ for $r_{12}\rightarrow\infty$ as opposed to an exponential decay for the Slater-type correlation factor in Eq.~\ref{eq:STGCorrFac}
commonly used in F12 theories. From consideration of the correct van der Waals description of a minimal basis helium dimer, the same long range $1/r_{12}$
form was deduced in Ref.~\onlinecite{Ten-no:JCP121-117}. However, since this long-range part of the correlation function is continuous and 
able to be captured in single reference theories by basis functions of angular momentum of $L_{\mathrm{occ}}+1$, it was not deemed necessary there to 
include this asymptotic behavior in the form of the correlation factor. In this paper, we will show clear improvements from the Yukawa-Coulomb correlation
factor in the case of the UEG where the correlation is isotropic, however it remains to be seen if any advantages are transferable to {\em ab initio} solid
state or extended molecular systems, where the longer range behavior in the geminals may be projected out by the determinantal expansion in the presence of
significant inhomogeneity in the potential.

Although we use simple perturbative arguments to motivate correlation functions of the uniform electron gas -- the prototypical
example of a metallic system where simple perturbation theory will fail -- it should be noted that for a two electron system the model is highly
insulating and metallic behavior and divergent results only arise on approach to the thermodynamic limit~\cite{Shepherd2013}. In addition, 
this long-range $1/r_{12}$ tail for the pair correlation function can also be motivated from the random phase 
approximation in this thermodynamic limit, where Gaskell\cite{Gaskell61,Ceperley78}
found the {\em exact} long-range behavior of the uniform electron gas to be 
\begin{equation}
\lim_{r_{12} \rightarrow \infty} u(r_{12}) \propto r_{12}^{-(D-1)/2}    ,
\end{equation}
where $D$ is the dimension of the model, and $e^{-u(r_{12})}$ then gives the exact solution to the two-body Schr{\"o}dinger equation. This gives
confirmation that the form of the correlation factor given in Eq.~\ref{eq:yccorrf} is {\em exactly} correct for both long and short distances in the
two-body correlation, although not necessarily between. This is also true in the strongly correlated regime, although there higher body effects
are obviously increasingly important. This knowledge has informed
the choice of Jastrow factors within the quantum Monte Carlo community\cite{Uschmajew2012}, whose simplest functional form of
\begin{equation}
u(r_{12}) = e^{-\frac{r_{12}}{2(1+br_{12})}}   ,
\end{equation}
also has the correct long-ranged $1/r_{12}$ behavior, and is used as standard for two-body correlation in both 
molecular and extended systems\cite{Handy69,Schmidt1990,Ceperley84,Umrigar08}. These Jastrow factors, which can be constructed to have
increasing numbers of variational parameters, additionally in higher particle number coordinates\cite{Hoggan2013}
capture all correlation effects of variational Monte Carlo methods\cite{Foulkes2001}. 

\section{\label{sec:results}Results}

This section discusses MP2 and MP2-F12 results of the
finite simulation cell uniform electron gas model.
Section~\ref{sec:mp2} recapitulates the well-known basis set extrapolation procedures
used in MP2 theory to obtain accurate complete basis set limit reference energies.
Section~\ref{sec:mp2f12comp} investigates the convergence of the
MP2-F12 correlation energy with respect to the employed computational parameters such as
the size of the CABS space, the variational parameter $\gamma$ used in the correlation
factors and the orbital basis set.
Having established CABS convergence, section~\ref{sec:optgam} examines the variation in the optimal parameter $\gamma$
governing the extent of the correlation hole, as the electron density of the system is changed.
Section~\ref{sec:optpair} explores the potential benefit of a pairwise optimization of the correlation factor
in order to accelerate the correlation energy convergence with respect to the
employed basis set even further.
Finally, section~\ref{sec:nonpar} investigates the relative accuracy in finite basis
set MP2 and MP2-F12 correlation energies as a function of the electron density.

\subsection{\label{sec:mp2}Basis set convergence in MP2 theory}

\begin{figure}
\includegraphics[width=8.0cm,clip=true]{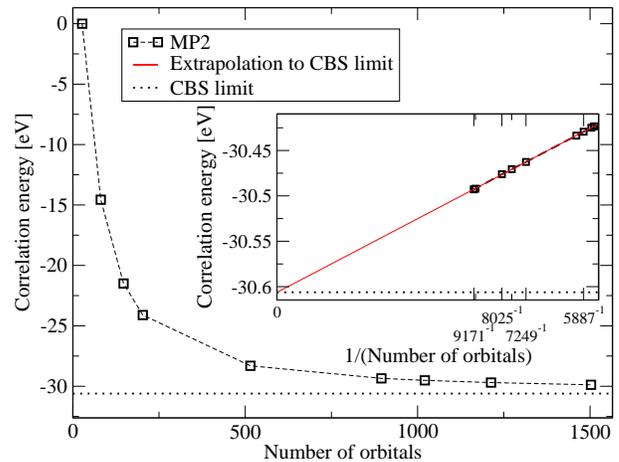}
\caption{\label{fig:MP2_rs5_conv} 
Convergence of the MP2 correlation energy for the 54 electron UEG simulation cell ($r_{\rm S}=5.0$~bohr)
with respect to the employed number of orbitals $M$.
The inset shows that the correlation energy behaves as $1/M$ using a very large orbital basis set, which
allows for the extrapolation to the complete basis set limit.}
\end{figure}

Accurate complete basis set limit MP2 correlation energies are an indispensable prerequisite
for the investigation of the quality of our MP2-F12 results.
To this end we outline the calculation of the MP2 complete basis set limit
energies below.
Figure~\ref{fig:MP2_rs5_conv} shows the convergence of the MP2 correlation energy
with respect to the employed basis set for 54 electrons in a box at a density corresponding to $r_{\rm s}=5.0$~bohr, a typical electron density for e.g. potassium metal.
As derived and discussed thoroughly in Ref.~\onlinecite{Shepherd2012_3}, the MP2 correlation energy
converges only as $1/M$ to the complete basis set limit, where $M$ corresponds to the number of plane waves.
This rate of convergence results directly from the convergence of the first-order cusp condition 
by the wavefunction.
In the present work, we employ this functional behavior ($1/M$) to extrapolate to the complete
basis set limit ($M\rightarrow \infty$).
The extrapolations were carried out using several MP2 energies obtained for orbital cutoffs
yielding 5887 to 9171 orbitals. 
The inset in figure~\ref{fig:MP2_rs5_conv} confirms that
the MP2 correlation energies for these basis sets converge as $1/M$ to the complete basis set limit.
In the following we will employ extrapolated complete basis set limit
energies as reliable comparisons for MP2-F12 results.

\subsection{\label{sec:mp2f12comp}Computational parameters in MP2-F12 theory}

\subsubsection{CABS convergence}

\begin{figure}
\includegraphics[width=8.0cm,clip=true]{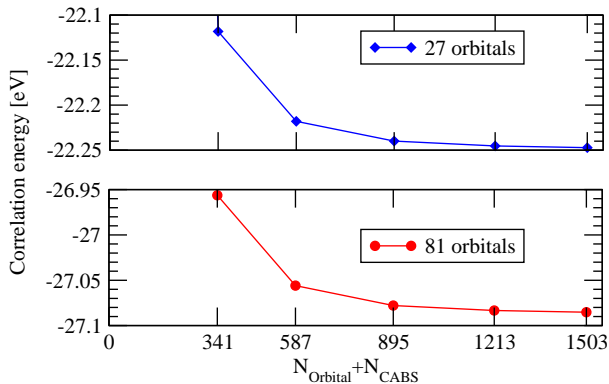}
\caption{\label{fig:ncabs_conv} 
Convergence of the MP2-F12 correlation energy for the 54 electron UEG simulation cell ($r_{\rm s}=5$~bohr)
with respect to the employed number of basis functions in the orbital basis and the CABS.
A Slater-type correlation factor and $\gamma=0.67$~\AA$^{-1}$ was used.
The number of basis functions
in the orbital space was fixed to 27 (upper panel) and 81 (lower panel). The CABS convergence rate does
not change as the number of virtual orbitals is increased.
}
\end{figure}

To avoid the explicit evaluation of three- and four-electron integrals in F12 calculations,
a complementary auxiliary basis set (CABS) is introduced\cite{Valeev:CPL395-190}. 
By insertion of the resolution of identity (RI),
many-electron integrals can be replaced with products of two electron integrals
contracted over the union of the orbital basis and the CABS
(\textit{e.g.} $\langle mnl | f_{12}f_{23} | lji \rangle \rightarrow \sum_P\langle mn | f_{12} | lP \rangle
\langle Pl | f_{12} | ji \rangle $).
In this work, the CABS space is trivially constructed as a set of higher momentum plane waves to those
in the orbital basis, and is therefore automatically orthogonal.
Figure~\ref{fig:ncabs_conv} demonstrates a rapid convergence of the MP2-F12 energy with respect
to the number of CABS orbitals used in the RI, and crucially, the rate of this convergence is independent of the 
orbital basis size. 
This is because the occupied orbitals do not include components of higher momentum as the orbital basis 
increases, and therefore the quality of the RI for a fixed number of electrons is independent of the size of 
the virtual basis, and only depends on the size of the complete basis set (the union of OBS and CABS).
In addition, we note that the MP2-F12 energy changes by less than
50~meV if the number of basis functions in the RI increases from 587 to 1503. As indicated in section~\ref{sec:introduction}, 
due to conservation of momentum, the RI for three-electron integrals in the uniform electron gas can in fact be saturated with a single function, obviating 
the need for a full RI in these cases. However, in order to maintain generality, this approach will not be considered further here.

This invariance with respect to orbital basis size will not be strictly true 
for {\em ab initio} systems, and so the question of convergence with respect to the auxiliary
basis will need to be readdressed at a later date. In addition, even for the uniform electron gas, the convergence will change
with number of electrons, as the formal requirement for saturation of the auxiliary basis for three-electron integrals 
includes plane waves with momenta $3 \times k_{\rm occ}$, 
where $k_{\rm occ}$ is the maximum momenta of the occupied orbitals\cite{Klopper87,Kutzelnigg:JCP94-1985}. 
However, errors may be sufficiently small such that the RI basis can be 
truncated well before this limit, and the computational cost for increasing the basis is only $\mathcal{O}[M^2]$. 
This issue will be returned to in the context of {\em ab initio} systems at a later date.

\subsubsection{$\gamma$ optimization}
\begin{figure}
\includegraphics[width=8.0cm,clip=true]{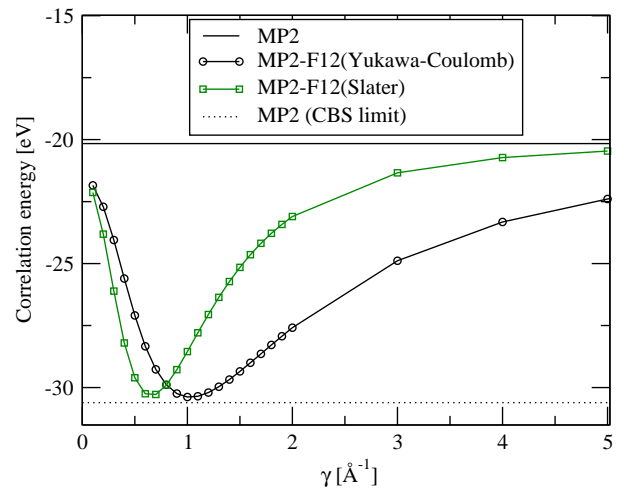}
\caption{\label{fig:optgamma}
The variation in the MP2-F12 energy with respect to the $\gamma$ parameter 
in the Slater- and Yukawa-Coulomb correlation factors, with the optimal
$\gamma$ giving the lowest MP2-F12 energy.
We employed 123 orbitals with 724 CABS basis functions for the 54 electron system at a
density of $r_{\rm s}=5$~bohr.
}
\end{figure}

As discussed in Section~\ref{sec:theory}, the $f_{12}^{\rm STG}$ and $f_{12}^{\rm YC}$ correlation factors
depend on the parameter $\gamma$ that 
describes how quickly the correlation factor decays to zero
with increasing inter-electronic distance, modeling the physical extent of the correlation hole.
The Hylleraas energy functional Eq.~\ref{eq:hylleraas} is variational and allows optimization
of $\gamma$ through energy minimization.
Figure~\ref{fig:optgamma} shows the dependence of the MP2-F12 energy
on $\gamma$ for the Slater and Yukawa-Coulomb correlation factors.
The Yukawa-Coulomb and Slater-type correlation factors minimize the MP2-F12 correlation energies
when $\gamma=1.04$~\AA$^{-1}$ and $\gamma=0.67$~\AA$^{-1}$, respectively.

It is instructive to compare the behavior of the two correlation factors by contrasting their series expansion about
$r_{12}=0$, which gives
\begin{align}
- \frac{f_{12}^{\rm STG}(r_{12})}{\gamma}=- \frac{1}{\gamma}+r_{12}-\frac{\gamma r_{12}^2}{2} + \mathcal{O}(r_{12}^3) \label{eq:taylorstg} \\
- \frac{f_{12}^{\rm YC}(r_{12})}{\gamma}= - \frac{2}{\gamma}+r_{12}-\frac{\gamma r_{12}^2}{3} + \mathcal{O}(r_{12}^3) \label{eq:taylorycc}.
\end{align}
The zeroth-order terms on the right hand side of the above equations are constant. Constant shifts in the correlation factors
are, however, always removed by the projector $\hat{Q}_{12}$ defined in Eq.~(\ref{eq:proj}) and yield
no contribution to the MP2-F12 correlation energy.
The first-order terms agree in both correlation factors, and are linear as required by the first-order cusp condition.
Inserting the optimized $\gamma$'s to calculate the coefficients for the second-order terms in $r_{12}$
from Eqs.~(\ref{eq:taylorstg}) and (\ref{eq:taylorycc}) yield
$0.347$~\AA$^{-1}$ and $0.335$~\AA$^{-1}$ for the Yukawa-Coulomb and Slater-type correlation factor respectively.
This comparison shows that both correlation factors give in fact very similar results at the cusp position for the present system.
However, the Yukawa-Coulomb correlation factor yields an improved minimum energy for the system, which is lower by approximately 100~meV 
compared to the Slater-type correlation factor, indicating its superior suitability for the system as expected.

We also note that the MP2-F12 energy becomes identical for both correlation factors in the limits $\gamma\rightarrow\infty$
and $\gamma\rightarrow 0$.
For $\gamma\rightarrow\infty$, the MP2-F12 energy converges to the conventional MP2 energy in the respective orbital
basis set. In the limit $\gamma\rightarrow0$, both correlation factors become $r_{12}$. As such, the latter limit corresponds to the
MP2-R12 correlation energy.

\subsubsection{Basis set convergence}
\begin{figure}
\includegraphics[width=8.0cm,clip=true]{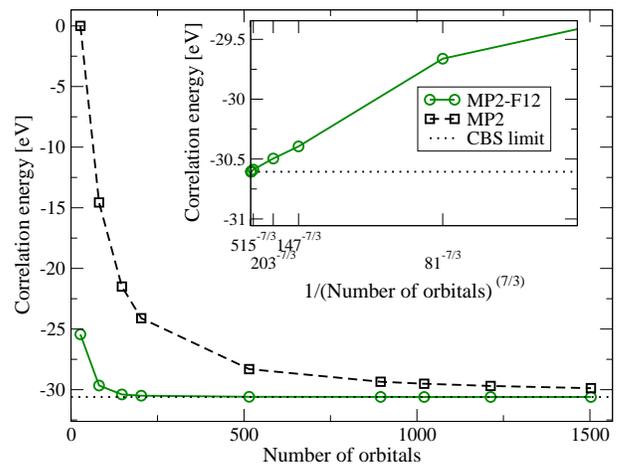}
\caption{\label{fig:MP2F12_rs5_conv} 
Convergence of the MP2 and MP2-F12 correlation energies for the 54 electron UEG simulation cell ($r_{\rm S}=5.0$~bohr)
with respect to the employed number of orbitals $M$ using the optimum $\gamma$ (see figure~\ref{fig:optgamma}).
}
\end{figure}

As a further test of the quality of the MP2-F12 we consider the convergence 
with respect to the orbital basis using the Slater-type
correlation factor, and compare to the extrapolated CBS limit results outlined in Section~\ref{sec:mp2}
and in Ref.~\onlinecite{Shepherd2012_3}.
Figure~\ref{fig:MP2F12_rs5_conv} confirms that the correct CBS limit correlation energy (30.61~eV) is recovered in the large
basis limit of our MP2-F12 implementation. As anticipated, we find that the rate of convergence
for the MP2-F12 results is greatly improved compared to conventional MP2 theory.
The inset in figure~\ref{fig:MP2F12_rs5_conv} shows that the MP2-F12 correlation energy converges
approximately as $1/M^{\frac{7}{3}}$, significantly faster than the $1/M$ convergence of MP2. This 
can be rationalized from the optimal convergence of a principal expansion of the wavefunction with
terms linear in $r_{12}$, which can be shown to be $(L+1)^{-7}$ where $L$ is the largest momentum
in the expansion\cite{Kutzelnigg:JCP94-1985}.

\begin{figure}
\includegraphics[width=8.0cm,clip=true]{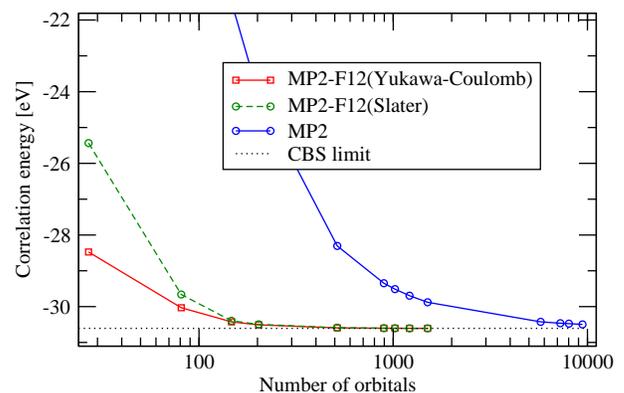}
\caption{\label{fig:MP2F12_conv_both}
Convergence of the MP2-F12 correlation energy for
the Slater- and Yukawa-Coulomb correlation factors at $r_{\rm s}=5.0$~bohr
with respect to the employed number of orbitals compared
to the MP2 energy and the CBS limit. It can be seen that MP2-F12 converges far quicker than MP2.
}
\end{figure}

Figure~\ref{fig:MP2F12_conv_both} shows the convergence of the MP2 and MP2-F12
correlation energies with respect to the employed basis set
for the Slater-type and Yukawa-Coulomb correlation factors.
We stress that a logarithmic scale is used on the horizontal axis.
As expected, both correlation factors converge to
the correct CBS limit. Furthermore the Yukawa-Coulomb correlation factor exhibits
a slightly faster rate of convergence indicating that the $1/r$ decay of
of $f^{\rm YC}$ captures longer-ranged, important correlation
effects that are neglected by the exponentially decaying Slater-type correlation factor.
The results shown in figure~\ref{fig:MP2F12_conv_both} suggest that MP2-F12 allows for a
reduction of the size of the orbital basis by approximately an order of magnitude, although often more.
Even though more investigation is required, and this factor will certainly not be fixed for different systems, 
this suggests savings in the orbital space could be on the whole larger than those 
generally achieved for molecular systems within a Gaussian orbital basis.

\subsection{\label{sec:optgam}Variation of $\gamma_{\rm opt}$ with electron density}
\begin{figure}
\includegraphics[width=8.0cm,clip=true]{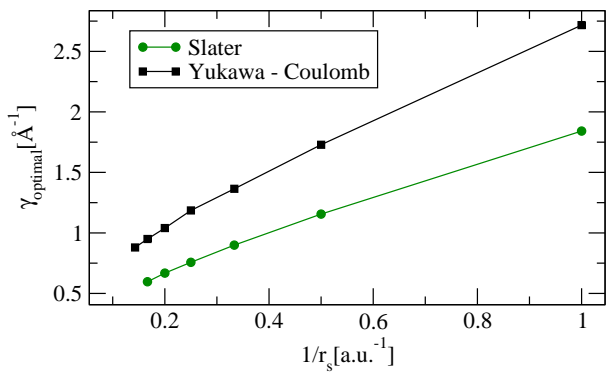}
\caption{\label{fig:gamma_vs_rs} 
Change in optimal $\gamma$ parameter for the Slater- and Yukawa-Coulomb correlation factors as
the density ($\propto r_{\rm s}$) of the electron gas is varied, for the 54 electron simulation cell.
}
\end{figure}

The physical extent of the correlation hole will change with the density of electrons, which in the electron gas
model we are considering is inversely proportional to the $r_{\rm s}$ parameter. Therefore, we expect the optimal
$\gamma$ for the correlation factors to increase for higher densities. This is indeed observed, as can be seen in
figure~\ref{fig:gamma_vs_rs}, which shows an approximately linear relationship between the optimal $\gamma$ and the
electron density ($1/r_{\rm s}$) for both correlation factors.
This linear relationship allows for the determination of an approximately optimal $\gamma$ in advance of any
calculation, without the need for an explicit optimization of the parameter with respect to the MP2-F12 energy.

\subsection{\label{sec:optpair}Pairwise $\gamma$ optimization}

\begin{figure}
\includegraphics[width=8.0cm,clip=true]{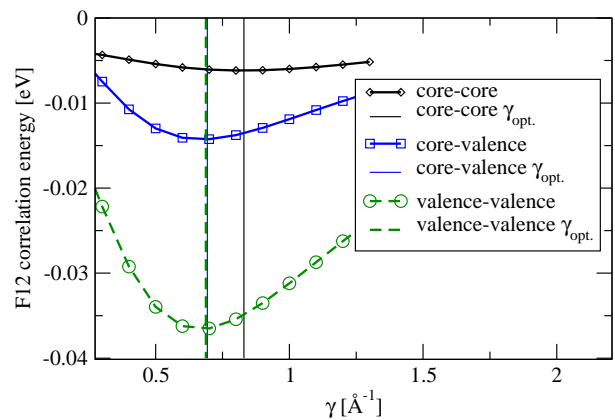}
\caption{\label{fig:gammasweep_pairwise} 
Pairwise optimization of $\gamma$ used for the Slater-type correlation factor
shown for a core-core, core-valence and valence-valence electron pair at $r_{\rm s}=5.0$~bohr.
}
\end{figure}

We now seek to investigate the potential improvement in the basis set convergence rate
of MP2-F12 by optimizing the correlation factor for each pair of electrons.
The correlation factor in MP2-F12 theory is known to depend on the orbital eigenvalues\cite{Tew08} 
and indeed our derivation of $f^{\rm YC}_{12}$ for doubly occupied pairs also reveals a
dependence of $\gamma$ on $\epsilon_i$, although this dependence is weak since it is partially canceled by $k_i^2$.
Since MP2 theory is an independent electron pair approximation one is free to use
different correlation factors for each electron pair.
Figure~\ref{fig:gammasweep_pairwise} shows the F12 correlation energy contributions
as a function of $\gamma$ for three different classes of electron pairs:
(i) a core-core electron pair, (ii) a core-valence electron pair and (iii) a valence-valence electron pair,
as defined by the kinetic energy of the electrons and their plane wave momenta, rather than their density 
since all plane waves have a uniform density across the computational cell.
Core and valence orbitals correspond to the plane wave orbitals with zero and the
highest possible kinetic energy for the present 54 electrons in a cubic box system.

The energy contributions are variational with respect to $\gamma$ and the respective minima are depicted by
vertical lines. The optimal $\gamma$ is found to be larger for core-core electron correlation than for
valence-valence and core-valence electron correlation. As such it would seem beneficial to employ
pairwise-optimized correlation factors. However, the contribution of the core-core correlation energy
in the present system is small compared to the contribution of the valence-valence electron pair energy.
Furthermore the additional correlation energy gained by the optimized correlation factor for the core-core electron
pair is almost negligible.
To this end we conclude that a pairwise optimization of the electron correlation factor is not a particularly worthwhile pursuit for
the uniform electron gas.
Furthermore this observation indicates that the remaining errors in the finite-basis MP2-F12 calculations using
optimized $\gamma$ values arise from the violation of higher-order cusp conditions.

\subsection{\label{sec:nonpar}Relative errors in plane wave MP2-F12 theory}

Although a rapid convergence of the absolute correlation
energy with respect to the employed basis set is advantageous for the study of real solid state systems,
it can be equally important that the \emph{rate of convergence} does not change significantly in the investigated coordinate space.
The latter allows for the calculation of properties such as lattice constants, bond lengths or reaction energies
in the complete basis set limit without having to converge the underlying absolute correlation energies, since absolute errors
are relatively constant, and therefore a cancellation of these errors yield accurate energy differences.
In the present system the errors in the MP2(-F12) correlation energies for a range of electron densities
with respect to a fixed basis set size, provides a good test case to investigate the issues described above.

\begin{figure}
\includegraphics[width=8.0cm,clip=true]{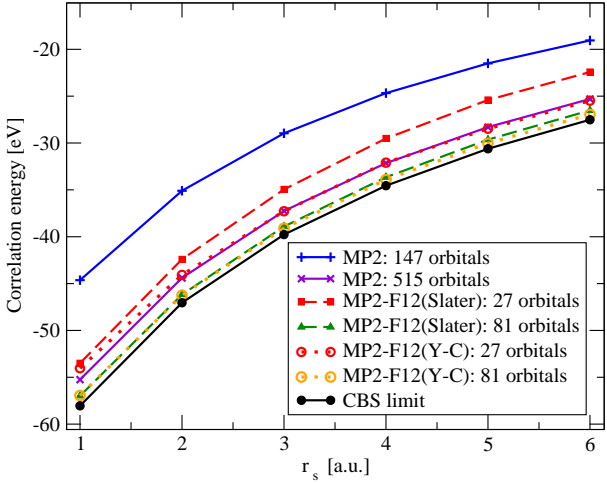}
\caption{\label{fig:MP2F12_vs_rs} 
MP2(-F12) correlation energies obtained using the optimal $\gamma$
over a range of densities given by the Wigner-Seitz radius $r_{\rm s}$ for 54 correlated electrons, compared to the CBS result.
}
\end{figure}

\begin{figure}
\includegraphics[width=8.0cm,clip=true]{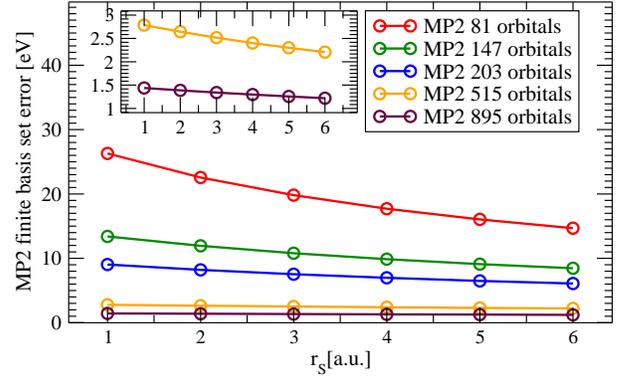}
\caption{\label{fig:absNPE_MP2.eps}
Basis set error of conventional MP2 correlation energies compared to the CBS limit, 
as a function of the Wigner-Seitz radius $r_{\rm s}$.}
\end{figure}

Figure~\ref{fig:MP2F12_vs_rs} shows the MP2(-F12) correlation energy as a function of $r_{\rm s}$ in the complete basis set limit
and for a range of finite basis sets. 
We find that the correlation energy increases in the limit of higher densities and that
finite as well as complete basis set limit results exhibit the same qualitative behavior
for increasing $r_{\rm s}$.
However, a more instructive plot is shown in figure~\ref{fig:absNPE_MP2.eps}, where the errors compared to the CBS result are given at each
electron density. This shows that the non-parallelity errors (the difference between the maximum and minimum basis set errors
over the electron densities considered) in finite basis conventional MP2 converge frustratingly slowly.
Employing 203 orbitals yields MP2 non-parallelity errors of approximately 2~eV over this density range, which roughly corresponds to
the range of realistic solid-state electron densities.

\begin{figure}
\includegraphics[width=8.0cm,clip=true]{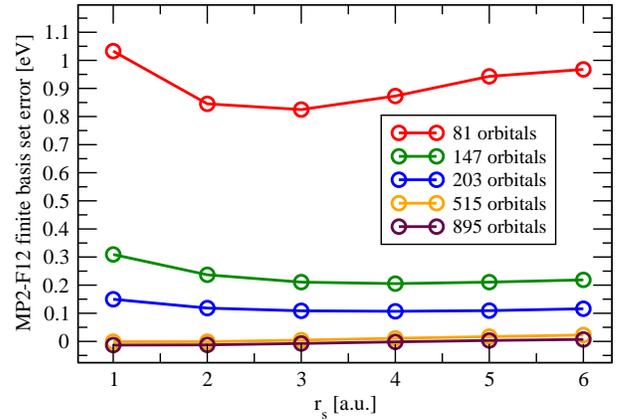}
\caption{\label{fig:F12_NPE_stgc.eps}
Basis set error of 54 electron MP2-F12 correlation energies compared to the CBS limit as a function of the Wigner-Seitz radius $r_{\rm s}$,
for the Slater-type geminal correlation factor.
From about 203 orbitals, the non-parallelity errors are almost negligible.
}
\end{figure}

\begin{figure}
\includegraphics[width=8.0cm,clip=true]{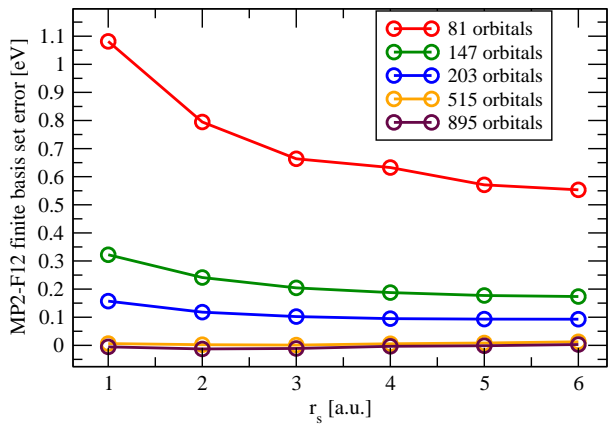}
\caption{\label{fig:F12_NPE_ycc.eps}
Basis set error of 54 electron MP2-F12 correlation energies compared to the CBS limit as a function of the Wigner-Seitz radius $r_{\rm s}$,
for the Yukawa-Coulomb geminal correlation factor.
From about 203 orbitals, the non-parallelity errors are almost negligible.
}
\end{figure}

In contrast to conventional MP2, MP2-F12
exhibits non-parallelity errors that converge much faster with respect to the basis set size.
Figures~\ref{fig:F12_NPE_stgc.eps} and \ref{fig:F12_NPE_ycc.eps} show the MP2-F12 errors compared
to the complete basis set limit for the same range of densities,
for the Slater-type correlation factor and Yukawa-Coulomb correlation factor respectively.
In contrast to the conventional MP2 result, 203 plane-wave orbitals suffice to obtain non-parallelity errors 
over this density range below 100~meV in the correlation energy, a reduction in the relative errors by over an order
of magnitude for the same basis size.

We note that 
the non-parallelity errors for all finite basis set results lead to a relative over-correlation
at lower densities, where longer ranged, non-dynamic correlation is more important, and therefore the
basis set convergence is seen to be faster.
The only exception to this observation is seen in the non-parallelity of the MP2-F12 energy
using the Slater-type correlation factor with 81 orbitals, as shown in figure~\ref{fig:F12_NPE_stgc.eps}.
In this case, the MP2-F12 basis set error exhibits a minimum at $r_{\rm s}=3$~a.u.
We believe that this indicates that the Slater-type
correlation factor is less efficient for lower densities where the long-range behavior
of the correlation factor is energetically more significant.

\section{\label{sec:con}Conclusions and Outlook}

In summary, we have shown that explicitly correlated MP2 theory can be
used in conjunction with a plane-wave basis set for three dimensional
fully periodic systems.
The combination of infinitely delocalized plane waves and a two-electron correlation
factor centered at the electron coalescence points spans a very efficient
and rapidly convergent basis set for the many-electron wavefunction expansion. This
allows for the accurate evaluation of the electronic correlation energy close to 
the complete basis set limit.
Our results for the uniform electron gas show that the reduction in the size of the employed one-electron
basis set is similar to the corresponding findings in Gaussian orbital based molecular systems, although
tentatively we suggest that the reduction could be even larger, perhaps due to the slower convergence
of the original plane wave basis compared to an optimized Gaussian-type orbital expansion.

We have introduced a novel correlation factor that is termed Yukawa-Coulomb correlation factor, which in
contrast to other employed correlation factors, is derived from analytic results for two electrons in a box.
The Yukawa-Coulomb correlation factor differs from the Slater-type correlation factor in the long range and
shows a faster rate of convergence with respect to the employed basis set.
We believe that this novel correlation factor may be useful for the study of solid state systems and potentially large molecules
with relatively isotropic interactions within explicitly correlated theories.

The change in the optimal variational parameter $\gamma_{\rm opt}$ was investigated
for a range of densities. We found that $\gamma_{\rm opt}$ increases linearly for larger electron densities, which indicates
that the correlation hole becomes more localized in this limit.
A close to optimal $\gamma$ can be determined solely from the density of the system and
the expectation is that even in {\em ab initio} systems, a $\gamma$ optimization will not always be necessary.

Furthermore we have investigated the pairwise optimization of the correlation factor for core-core, core-valence
and valence-valence electron pairs. Our findings show that although $\gamma_{\rm opt}$ for the core-core electron pairs
differs significantly from $\gamma_{\rm opt}$ for valence-valence electron pairs, the gain in the absolute correlation energy
using pairwise optimized correlation factors is negligible.
As such we believe that it is not beneficial to optimize the correlation factor for each
electron pair individually.

Finally we have studied the convergence of the non-parallelity error from the complete basis set limit
using MP2-F12 and MP2 for a range of densities and basis sizes.
This is expected to provide a good test case for the convergence of lattice constants and other energy differences in solid state systems
with respect to the employed basis set.
As expected the convergence of MP2-F12 clearly outperforms MP2 and also allows for a reduction by approximately an order of magnitude
in the employed basis set.

\begin{figure}
\includegraphics[width=8.0cm,clip=true]{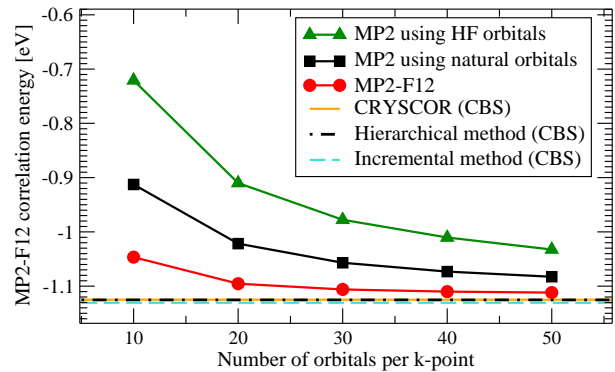}
\caption{\label{fig:LiH.eps}
Basis set convergence of the MP2 valence-only cohesive energy contribution to LiH.
Our calculations were done using a 4$\times$4$\times$4 $k$-mesh and norm-conserving pseudo-potentials
in the framework of the PAW method\cite{Kresse96}. The LiH unit cell volume
was set to 17.03~\AA$^3$. The MP2 calculations were done using HF and approximate natural orbitals\cite{Kresse2011}.
Our MP2 and MP2-F12 (using a STG) results converge to the same complete basis set limit results obtained
using local MP2 (CRYSCOR, from Ref.~\onlinecite{Schutz2011}), the hierarchical method (from Ref.~\cite{Manby2009})
and the incremental scheme (from Ref.~\cite{Stoll2012}).
}
\end{figure}

We hope that the findings of the present work
will translate both to alternative UEG models\cite{Gill2012,Gill2013}, {\em ab initio} systems, and to other 
explicitly correlated methods in the solid state such as
CCSD-F12\cite{Noga:JCP101-7738,Tew:PCCP9-1921,Tew:BOOK2010,Werner:BOOK2010,Knizia2007} or 
FCIQMC-F12\cite{BGKA2013,Booth2012,BTA2009,BoothC2,CBOA2012,CBA2010}, where the additional computational cost for
calculating the F12 contribution becomes negligible in comparison to these more expensive parent methods.
The application of the methods outlined in this work to real, {\it ab initio} solid state systems is
expected to significantly expand the scope of the whole range of quantum chemical wave function based 
methods. Figure~\ref{fig:LiH.eps} shows a preliminary application of the MP2-F12 implementation for
the LiH crystal confirming our findings for the uniform electron gas that explicitly correlated MP2 theory
allows for a substantial reduction in the basis set. We will expand on these results in a forthcoming paper.

\section{Acknowledgements}
The authors thank Jiri Klimes and Cyrus Umrigar for fruitful discussions.
A.G. acknowledges an APART-fellowship of the Austrian Academy of Sciences.
J.J.S. thanks EPSRC for funding.
A.A. acknowledges support from the EPSRC under grant number:  EP/J003867/1.
D.P.T. thanks the Royal Society for a Univerity Research Fellowship.
G.H.B. acknowledges funding from Trinity College, Cambridge.



\end{document}